\documentclass{llncs}

\usepackage{graphicx}
\usepackage{tikz}

\newcommand{\etal}{{et al.}}

\begin{document}
\frontmatter          % for the preliminaries
\pagestyle{headings}  % switches on printing of running heads

\mainmatter              % start of the contributions
\title{The NA62 RICH detector%
%: construction and performance%
}

\titlerunning{The NA62 RICH detector}  % abbreviated title (for running head)
\subtitle{construction and performance}

\author{Andrea Bizzeti\inst{1,2}\thanks{%
on behalf of the NA62 RICH Working Group: A.~Bizzeti,
 F.~Bucci, R.~Ciaranfi, E.~Iacopini, G.~Latino, M.~Lenti, R.~Volpe,
 G.~Anzivino, M.~Barbanera, E.~Imbergamo, R.~Lollini, 
 P.~Cenci, V.~Duk, M.~Pepe, M.~Piccini
%%% *****  EV.  ADD  INSTITUTES  ???   *******
}  % footnote
}  % author
\authorrunning{Andrea Bizzeti} % abbreviated author list (for running head)
\tocauthor{Andrea Bizzeti}
\institute{%
Department of Physics, Informatics and Mathematics,\\
 University of Modena and Reggio Emilia
\and
Istituto Nazionale di Fisica Nucleare -- Sezione di Firenze, Italy\\
E-mail: \email{andrea.bizzeti@fi.infn.it}
%,\\ WWW home page: \texttt{http://users/\homedir iekeland/web/welcome.html}
}

\maketitle              % typeset the title of the contribution

\begin{abstract}
%[*********** 70--150 words ***********]
%  
The RICH detector of the NA62 experiment at CERN SPS 
 is required to suppress $\mu^+$ contamination in
%  the $\pi^+$ sample
  $K^+ \to \pi^+ \nu \bar\nu$ candidate events
 by a factor at least 100 between 15 and 35~GeV/c momentum,
to measure the pion arrival time with $\sim 100$~ps resolution 
and to produce a trigger for a charged track.
It consists of a 17~m long tank
% (vessel) 
 filled with Neon gas at atmospheric pressure.
\v{C}erenkov light is reflected by a mosaic of 20 spherical mirrors
 placed at the downstream end of the vessel
and is collected by 1952 photomultipliers
% (PMTs)
 placed at the upstream end.
The construction of the detector will be described 
 and the performance reached during first runs will be discussed.
\end{abstract}

\section{The NA62 RICH detector}
The NA62 experiment\cite{ref:na62_proposal} 
 at CERN SPS North Area 
  has been designed to study charged kaon decays
 and particularly
to measure the branching ratio ($\approx 10^{-10}$) of
  the very rare decay $K^+ \to \pi^+ \nu \bar\nu$
 with a 10\% precision.
The NA62 experimental apparatus
is described in detail in~\cite{ref:na62_detector}.

%\section{The RICH detector}

The largest background to
 $K^+ \to \pi^+ \nu \bar\nu$ 
 comes from the $K^+ \to \mu^+ \nu_\mu$ decay,
which is 10 orders of magnitude more abundant.
This huge
 background is mainly suppressed using kinematical methods 
 and the very different response of calorimeters to muons
  and charged pions.
Another factor 100 in muon rejection is needed 
 in the momentum range between 15 and 35~GeV/c.
A dedicated \v{C}erenkov detector, the RICH, 
 has been designed and built for this purpose.
Neon gas at atmospheric pressure is used as radiator, 
 with refraction index $n = 1 + 62.8 \times 10^{-6}$ 
 at
a wavelength 
 $\lambda=300\mu$m,
 corresponding to a \v{C}erenkov threshold for charged pions 
  of 12.5~GeV/c.
Two full-length prototypes
 were built and tested with hadron beams
to study the performance
 of the proposed layout\cite{ref:rich_proto1,ref:rich_proto2}.

\begin{figure}
\includegraphics[width=\textwidth,keepaspectratio,clip,viewport=20 120 655 430]{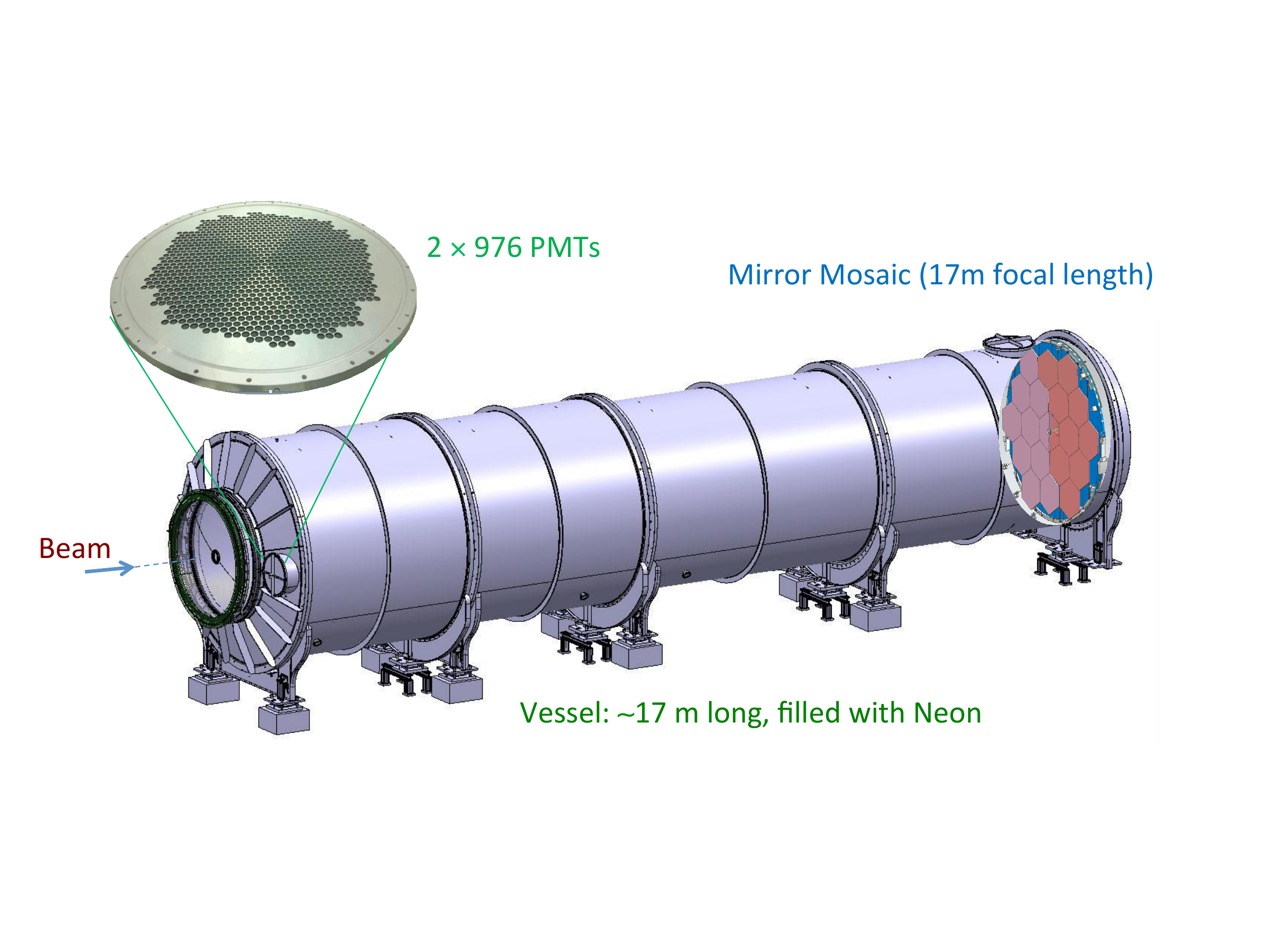}
\caption{Schematic view of the RICH detector.
The hadron beam enters from the left and travels throughout the length
 of the detector in an evacuated beam pipe.
A zoom of one of the two disks hosting the photomultipliers
 is shown on the left;
the mirror mosaic is made visible through the
 vessel
on the right.
From\cite{ref:na62_detector}.
% under the terms of the
% Creative Commons Attribution 3.0 License.
%(Fig.41).
%
}
\label{fig:rich}
\end{figure}

The RICH radiator container (``vessel'')
 is a 17~m long vacuum-proof steel tank,
 composed of 4 cylindrical sections of diameter
% ranging from
up
 to 4~m, closed by two thin aluminium windows
  to minimize the material budget crossed by particles.
A sketch of the RICH detector is shown in Fig.\ref{fig:rich}.
Fresh neon at atmospheric pressure
 are injected in the
200~m$^3$ vessel volume 
% vessel
  after it has been completely evacuated.
No purification or recirculation system is used.

% \subsubsection{The mirror mosaic.}

A mosaic of spherical mirrors
with 17.0~m focal length,
 18 of hexagonal shape (350~mm side)
 and two semi-hexagonal 
  with a circular opening
% to accomodate
 for the beam pipe,
is placed at the downstream end of the vessel
 to reflect and focus \v{Cerenkov} light
towards the two regions equipped with photomultipliers (PMTs)
at the upstream end of the vessel (see Fig.\ref{fig:rich}).

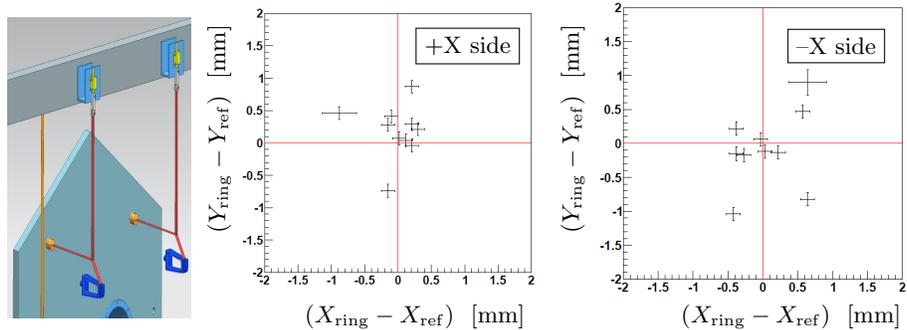
\begin{figure}
\mbox{
\begin{minipage}{0.2\textwidth}
\includegraphics[width=\textwidth,keepaspectratio,clip,viewport=70 0 440 607]{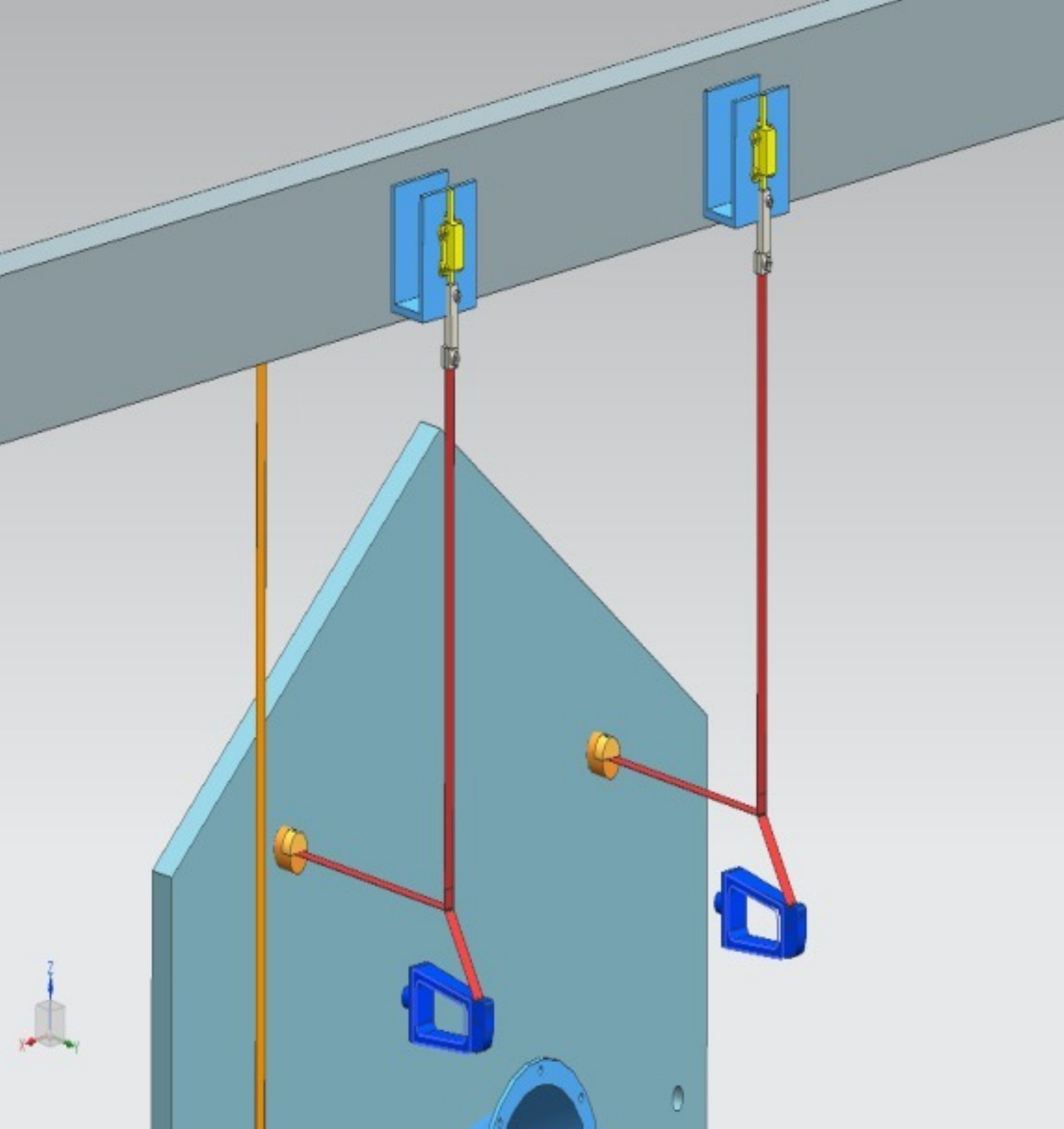}
\end{minipage}
\begin{minipage}{0.7\textwidth}
\begin{tikzpicture}
\node[anchor=south west,inner sep=0] at (1,1){%
\includegraphics[width=0.5\textwidth,keepaspectratio]{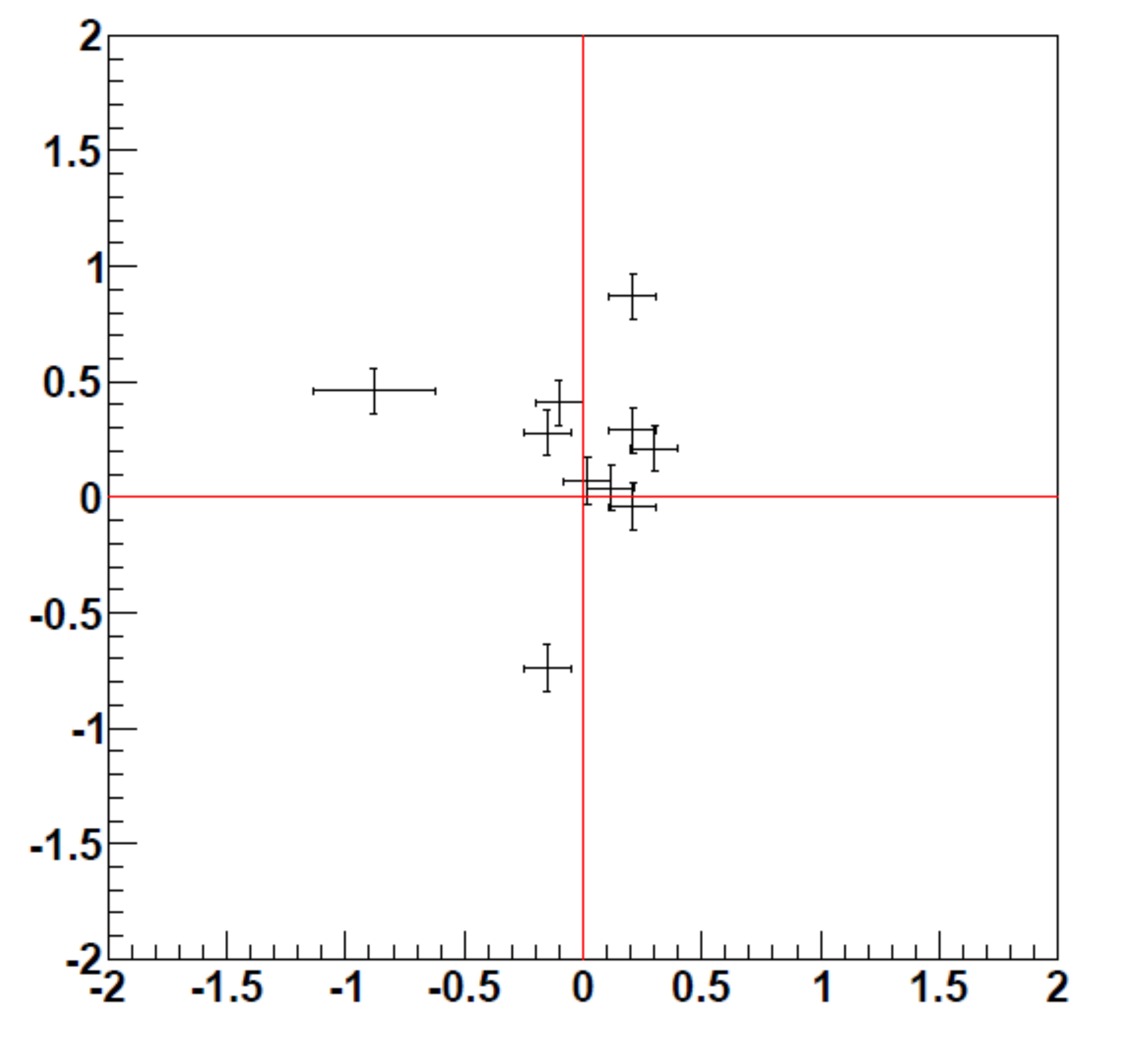}};
\node[anchor=south west,inner sep=0] at (5.8,1){%
\includegraphics[width=0.5\textwidth,keepaspectratio]{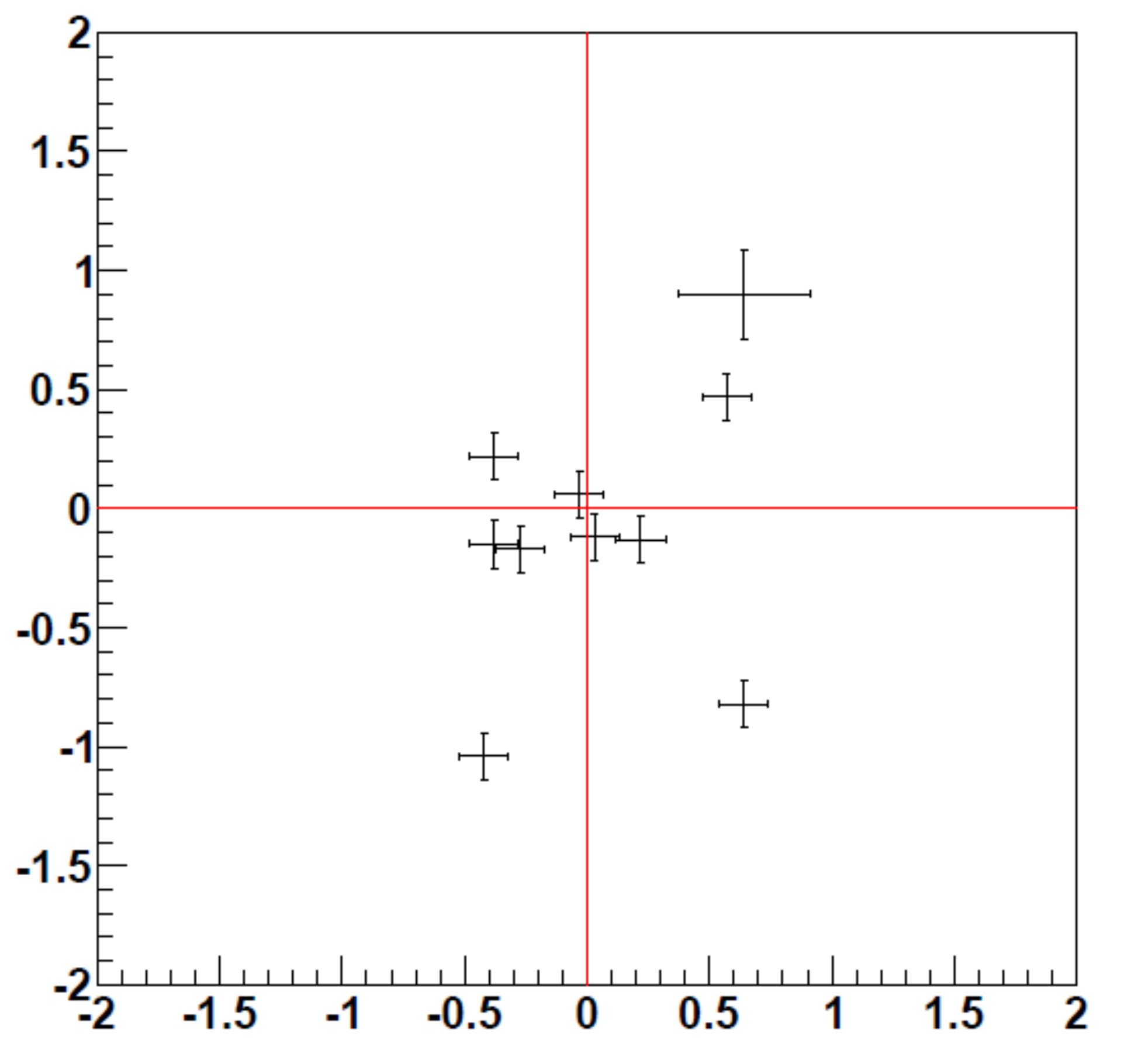}};

\node at (3.4,0.8) {$(X_\mathrm{ring}-X_\mathrm{ref})$ \ [mm]};
\node at (8.4,0.8) {$(X_\mathrm{ring}-X_\mathrm{ref})$ \ [mm]};
\node[rotate=90] at (0.8,3.3) {$(Y_\mathrm{ring}-Y_\mathrm{ref})$ \ [mm]};
\node[rotate=90] at (5.5,3.3) {$(Y_\mathrm{ring}-Y_\mathrm{ref})$ \ [mm]};

\node at (4.1,4.4) {\fbox{{\small +X side}}};

\node at (9,4.4) {\fbox{{\small --X side}}};

\end{tikzpicture}
\end{minipage}
}  % mbox

\caption{(left)
Scheme of the mirror orientation system:
 two ribbons connected to piezo-electric motors
  pull micrometrically the mirror
while a third vertical ribbon avoids on-axis rotation.
(centre and right) 
Position difference
 between the centre of the \v{C}erenkov ring 
 reconstructed by the RICH PMTs
and its expected position based on the track direction 
 reconstructed by the spectrometer,
  after tuning the mirrors orientation.
``+X side'' and ``--X side'' indicate the two locations of the PMTs,
 on the left and on the right of the beam pipe.
Each point represents a single mirror.
% 1~mm corresponds to $\sim 30\mu$rad in the mirror orientation.
}
\label{fig:mirrors}
\end{figure}

Each mirror is supported using a dowel inserted in 
 a 12~mm diameter
cylindrical 
  hole  drilled in the rear face of the mirror
 close to its barycentre.
Dowels are connected to the RICH vessel by means of 
 a vertical support panel, made of an aluminum honeycomb structure
to minimize the material budget.
Two thin aluminium ribbons,
each one
 pulled by a micrometric piezo-electric motor,
 keep the mirror in equilibrium and allow to modify its orientation.
A third vertical ribbon, without motor, 
 avoids on-axis rotation.
The mirror orientation
 is measured by comparing the position
 of the centre of the \v{C}erenkov ring 
 reconstructed by the RICH PMTs
with its expected position based on the track direction 
 reconstructed by the spectrometer
and can be finely tuned using piezomotors
 (see Fig.\ref{fig:mirrors}).
 
% \subsubsection{Photodetectors and electronics.}

Two arrays of 976 Hamamatsu R7400-U03 PMTs are located
 at the upstream end of the vessel,
 left and right of the beam pipe.
The PMTs have an 8~mm diameter active region 
 and are packed in an hexagonal lattice with 18~mm pixel size.
Each PMT has a bialkali cathode,
 sensitive between 185 and 650~nm wavelength
with about 20\% peak quantum efficiency at 420~nm.
Its 8-dynode system provides a gain of $1.5 \times 10^6$
 at 900~V supply voltage, with a time jitter of 280~ps FWHM.
PMTs are located in air outside the vessel
 and are separated from neon 
 by a quartz window;
 an aluminized mylar Winston cone\cite{ref:winston}
  is used to reflect incoming light 
   to the active area of each PMT.
The front-end electronics consists of 64 custom made boards,
  each of them equipped with four 8-channels
  Time-over-Threshold NINO discriminator chips\cite{ref:nino}.
The readout is provided by 4 TEL62 boards,
 each of them equipped with sixteen 32-channels HPTDC\cite{ref:hptdc};
 a fifth TEL62 board receives a multiplicity output
  (logic OR of the 8 channels)
  from each NINO discriminator
 and is used for triggering.
The time resolution of \v{C}erenkov rings
has been measured
by comparing
the average times of two subsets of the PMT signals,
resulting in $\sigma_t \mbox{(ring)} = 70$~ps.

\section{Particle identification}

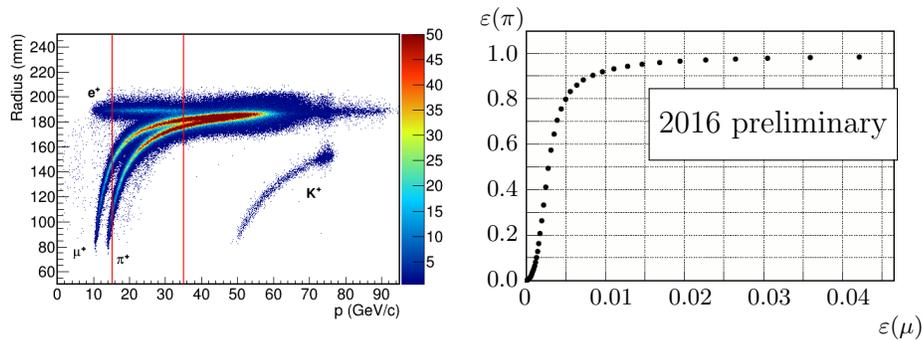
\begin{figure}

%\hspace*{-0.02\textwidth}
\mbox{
\begin{minipage}{0.48\textwidth}
\begin{tikzpicture}

\node[anchor=south west,inner sep=0] at (6.5,0.3){\includegraphics[width=\textwidth,clip,viewport=30 0 660 460]{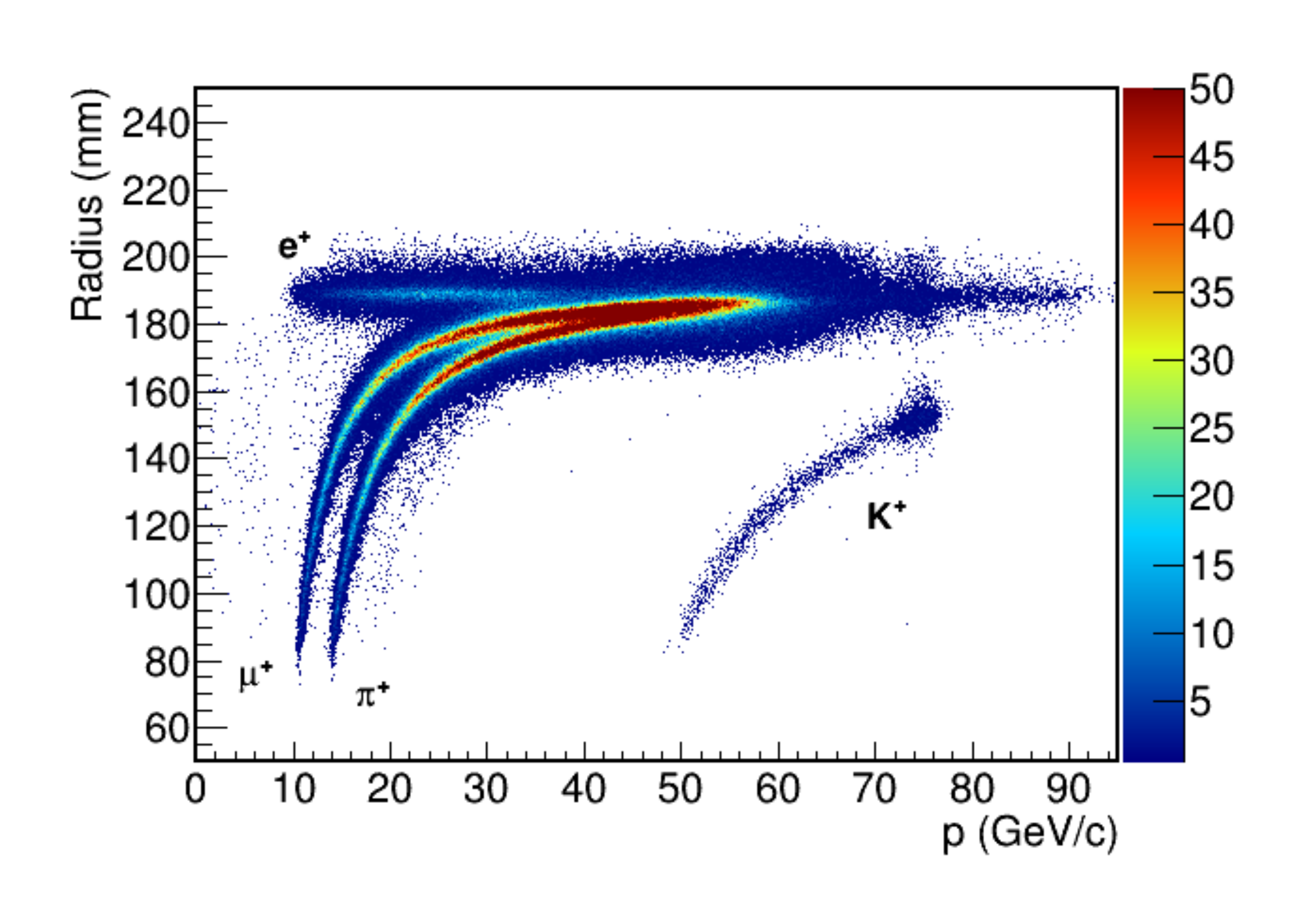}};

%\node[rotate=30] at (4.7,2.6) {{\underline{2015 data}}};

\draw[red] (7.92,0.97) -- (7.92,4.3);
\draw[red] (8.87,0.97) -- (8.87,4.3);

\end{tikzpicture}

\end{minipage}
\hspace{0.01\textwidth}
\begin{minipage}{0.45\textwidth}
\begin{tikzpicture}

\node[anchor=south west,inner sep=0] at (1,1){\includegraphics[width=0.9\textwidth,clip,viewport=61 41 600 410]{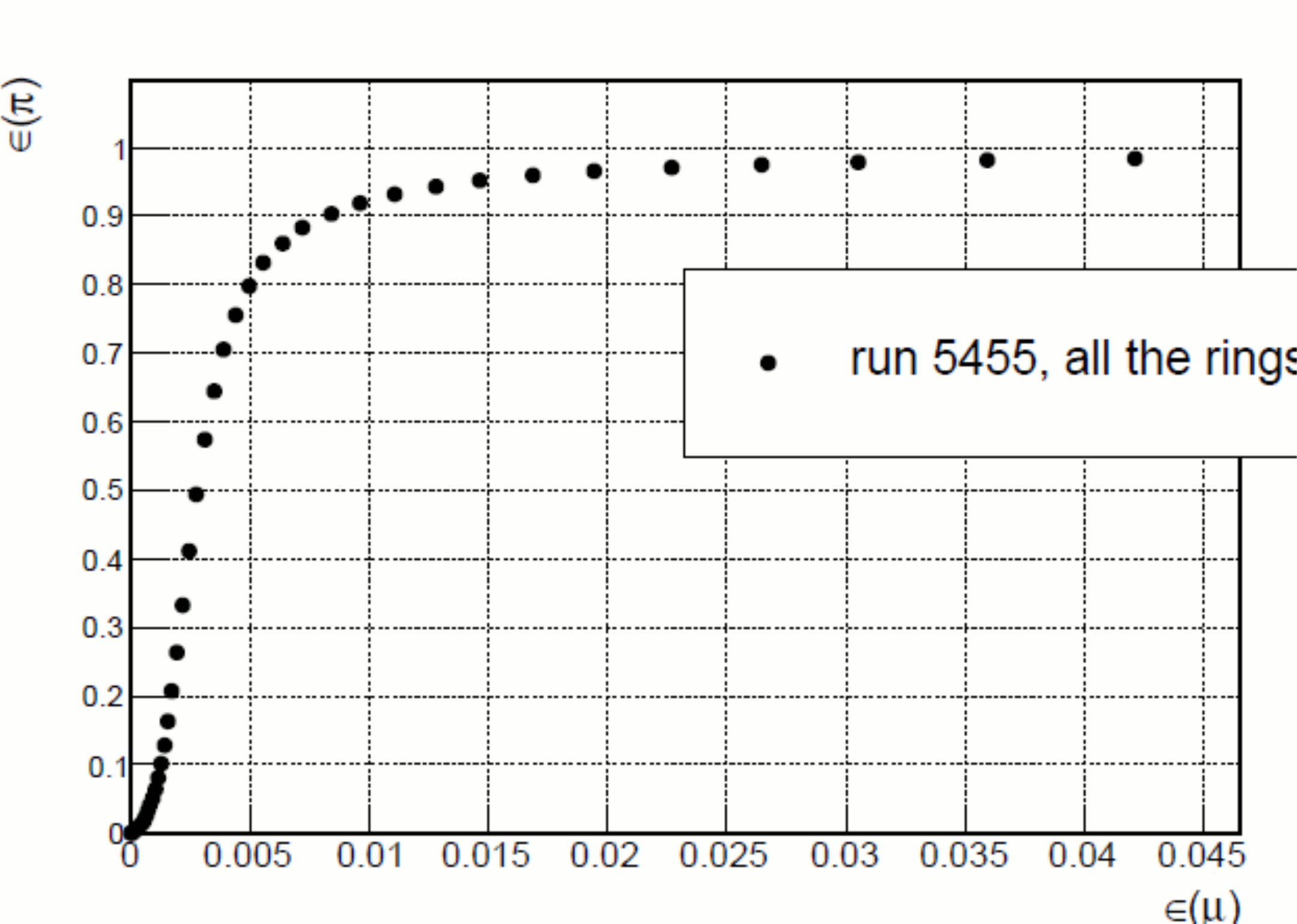}};

%\node at (3.5,6.0) {{$\pi^\pm$} \ identification efficiency vs.};
%\node at (3.7,5.5) {{$\mu^\pm$} \ mis-identification probability};

\node at (0.7,4.5) {{\small {$\varepsilon(\pi)$}}};

\node at (0.7,1.05) {{\footnotesize 0.0}};
\node at (0.7,1.65) {{\footnotesize 0.2}};
\node at (0.7,2.25) {{\footnotesize 0.4}};
\node at (0.7,2.85) {{\footnotesize 0.6}};
\node at (0.7,3.45) {{\footnotesize 0.8}};
\node at (0.7,4.05) {{\footnotesize 1.0}};

\node at (1,0.8) {{\footnotesize 0}};
\node at (2.15,0.8) {{\footnotesize 0.01}};
\node at (3.2,0.8) {{\footnotesize 0.02}};
\node at (4.25,0.8) {{\footnotesize 0.03}};
\node at (5.3,0.8) {{\footnotesize 0.04}};

\node at (6,0.4) {{\small {$\varepsilon(\mu)$}}};

% \draw[red] (1.9,1) -- (1.9,4.35);

\node at (4.3,3.1) {\large \fcolorbox{black}{white}{
\begin{minipage}{28truemm}
\vspace*{5pt}
\centerline{{2016 preliminary}}

%\node at (5,3.3) {\Large \fcolorbox{red}{white}{
%\begin{minipage}{24truemm}
%\vspace*{5pt}
%centerline{{2016 data}}

\vspace*{5pt}
\end{minipage}
}};

\end{tikzpicture}
\end{minipage}
}  % mbox

\caption{%Particle identification with the RICH
%RICH detector performance:
(left) \v{C}erenkov ring radius versus particle momentum.
Vertical lines delimit the momentum fiducial region 15--35~GeV/c.
Electrons, muons, charged pions and scattered beam kaons
 are clearly visible.
Particles with momentum higher than 75~GeV/c correspond to halo muons.
(right) Pion identification efficiency
 versus muon mis-identification probability.
}
\label{fig:pid}
\end{figure}

In order to assess the RICH performance,
the \v{C}erenkov ring radius (which depends on particle velocity)
 measured by the RICH
  is related to the track momentum
 measured by the magnetic spectrometer.
Figure~\ref{fig:pid}(left)
 shows a clear separation between different particles 
 in the momentum range 15--35~GeV/c.
Pion-muon separation is achived by cutting on the particle mass,
 calculated from the measured values of the particle velocity
  (from the \v{C}erenkov ring radius)
 and momentum.
The charged pion identification efficiency $\varepsilon_\pi$
 and muon mis-identification probability $\varepsilon_\mu$
are plotted in Fig.~\ref{fig:pid}(right)
 for several values of the mass cut.
%  smaller than 1\% is measured 
At $\varepsilon_\pi=90\%$
% pion identification efficiency
 the muon mis-identification probability
 is $\varepsilon_\mu \simeq 1\%$.
% (see Fig.\ref{fig:pid}(right)).

\section{Conclusions}

The NA62 RICH detector 
 was installed in 2014 and commissioned in autumn 2014 and 2015;
it is fully operational since the 2016 run.
First performance studies with collected data
 show that the RICH
  fulfilled the expectations, 
achieving a time resolution of 70~ps
 and a factor $\sim 100$ in muon suppression.

\section*{Acknowledgements}

The construction of the RICH detector would not have been possible 
 without the enthusiastic work of many technicians from University and INFN
  of Perugia and Firenze, the staff of CERN laboratory,
 the collaboration with Vito Carassiti from INFN Ferrara.
A special thank to the NA62 collaboration 
 for the full dedication to the construction, commissioning and running
  of the experiment.


\end{document}